\documentclass[conference]{IEEEtran}
\IEEEoverridecommandlockouts
\usepackage{cite}
\usepackage{amsmath,amssymb,amsfonts}
\usepackage{algorithmic}
\usepackage{longtable}
\usepackage{tabularx}
\usepackage{graphicx}
\usepackage{textcomp}
\usepackage{xcolor}
\usepackage{array}
\usepackage{tabularx}
\usepackage{multirow}
\usepackage{booktabs}
\usepackage{array}
\usepackage{tikz}
\usepackage{pdflscape}
\usetikzlibrary{positioning}
\newcolumntype{Y}{>{\raggedright\arraybackslash}X}

\usepackage{makecell}
\def\BibTeX{{\rm B\kern-.05em{\sc i\kern-.025em b}\kern-.08em
    T\kern-.1667em\lower.7ex\hbox{E}\kern-.125emX}}
\begin{document}


\title{A Blueprint for AI-Driven Software Quality: Integrating LLMs with Established Standards}

\author{\IEEEauthorblockN{Avinash Patil}
    \IEEEauthorblockA
    {
    \textit{Juniper Networks Inc.}\\
    avinashpatil@ieee.org\\
    ORCID: 0009-0002-6004-370X
    }
}

\maketitle

\begin{abstract}
Software Quality Assurance (SQA) is critical for delivering reliable, secure, and efficient software products. The Software Quality Assurance Process aims to provide assurance that work products and processes comply with predefined provisions and plans. Recent advancements in Large Language Models (LLMs) present new opportunities to enhance existing SQA processes by automating tasks like requirement analysis, code review, test generation, and compliance checks. Simultaneously, established standards such as ISO/IEC 12207, ISO/IEC 25010, ISO/IEC 5055, ISO 9001/ISO/IEC 90003, CMMI, and TMM provide structured frameworks for ensuring robust quality practices. This paper surveys the intersection of LLM-based SQA methods and these recognized standards, highlighting how AI-driven solutions can augment traditional approaches while maintaining compliance and process maturity. We first review the foundational software quality standards and the technical 
fundamentals of LLMs in software engineering. Next, we explore various LLM-based SQA applications, including requirement validation, defect detection, test generation, and documentation maintenance. 
We then map these applications to key software quality frameworks, illustrating how LLMs can address specific requirements and metrics within each standard. Empirical case studies and open-source initiatives demonstrate the practical viability of these methods. At the same time, discussions on challenges (e.g., data privacy, model bias, explainability) underscore the need for deliberate governance and auditing. Finally, we propose future directions encompassing adaptive learning, privacy-focused deployments, multimodal analysis, and evolving standards for AI-driven software quality. By uniting insights 
from academic research, industry best practices, and established quality frameworks, we provide a comprehensive blueprint for integrating LLMs into SQA in a trustworthy, efficient, and standards-aligned manner.
\end{abstract}


\section{Introduction}
\label{introduction}

Software systems continue to grow in complexity and scope, with modern applications often integrating numerous services and components \cite{Cito2018}. As a result, Software Quality Assurance (SQA) activities must adapt to ensure reliability, security, and maintainability across distributed architectures. A key challenge arises from the increasing velocity of software releases, necessitating more automated and intelligent methods to maintain high-quality standards.

Large Language Models (LLMs) have emerged as a powerful asset in addressing these challenges. By leveraging advanced neural architectures---particularly the transformer model introduced by Vaswani et al. \cite{Vaswani2017}---LLMs can learn contextual relationships from extensive natural language text and source code repositories. Recent works highlight their applicability to various software engineering tasks, including code completion and defect detection \cite{Chen2021, Feng2020}. For example, GitHub Copilot has demonstrated the potential to reduce coding effort by providing real-time suggestions, though questions remain regarding its accuracy and coverage \cite{Chen2021}.

The convergence of LLMs and SQA presents an opportunity to automate traditionally labor-intensive tasks, from identifying ambiguous requirements \cite{Poldrack2022} to generating comprehensive test suites \cite{Tian2018}. Although these approaches show promise, they must be systematically integrated into established frameworks and standards that guide quality assurance across industries \cite{patil2025next}. Standards such as ISO/IEC 12207, ISO/IEC 25010, ISO/IEC 5055, ISO 9001 (and 90003), CMMI, and the Test Maturity Model (TMM) have long provided robust structures for ensuring software quality \cite{Laplante2018}. As organizations explore AI-driven methods, aligning these techniques with known best practices is crucial for consistent, reliable outcomes \cite{jadon2025ethical}.

This survey examines how LLM-based techniques can enhance, extend, or streamline SQA activities within the context of recognized software quality standards. We begin with an overview of the major standards and then explore the fundamentals of large language models in software engineering. Subsequent sections delve into specific LLM-based approaches for requirements analysis, code review, testing, maintenance, and compliance. We present real-world case studies that illustrate the practical adoption of these methods, followed by an in-depth discussion of associated challenges, limitations, and risks. Finally, we conclude with prospective directions for research and policy, emphasizing the need for standards evolution to fully harness the potential of LLM-assisted SQA.

\section{Overview of Software Quality Standards}
\label{sec:quality-standards}

Software Quality Assurance (SQA) relies on established standards to guide the design, development, and maintenance of reliable, secure, and performant systems. This section provides an overview of key standards and models that underpin traditional SQA practices. In later sections, we will explore how large language model (LLM)- driven approaches can be aligned with or extended by these frameworks.

\subsection{ISO/IEC 12207 (Software Life Cycle Processes)}
\label{subsec:iso-12207}
ISO/IEC 12207 was published on 1 August 1995. It was the first International Standard to provide a comprehensive set of life cycle processes, activities, and tasks for software that is part of a larger system and for stand-alone software products and services.
ISO/IEC 12207 defines a comprehensive set of life cycle processes for software, spanning planning, development, operation, maintenance, and decommissioning \cite{ISOIEC12207}. It emphasizes a systematic and standardized approach throughout the software life cycle, mandating well-defined roles, responsibilities, and deliverables for each phase.  
\null\quad The standard underscores the importance of verification and validation (V\&V) processes, ensuring that products meet user requirements and intended purposes \cite{Laplante2018}. By integrating SQA tasks into these life cycle processes, ISO/IEC 12207 provides a robust framework for achieving software quality from concept definition to retirement.

\subsection{ISO/IEC 25010 (Systems and Software Quality Requirements and Evaluation -- SQuaRE)}
\label{subsec:iso-25010}
Published under the SQuaRE series, ISO/IEC 25010 presents a quality model that classifies software product quality characteristics into eight main categories, including functional suitability, performance efficiency, compatibility, usability, reliability, security, maintainability, and portability \cite{ISOIEC25010}.  
\null\quad The model aims to provide stakeholders with a shared understanding of essential quality attributes and how to measure them. Researchers have employed ISO/IEC 25010 as a basis for empirical evaluations of software products, revealing how different projects prioritize certain attributes (e.g., security over performance) depending on domain-specific needs \cite{Garousi2019}.

\subsection{ISO/IEC 5055 (Measuring Internal Software Quality)}
\label{subsec:iso-5055}
ISO/IEC 5055:2021 outlines a standard for detecting and measuring structural weaknesses in software systems by focusing on four critical factors: security, reliability, performance efficiency, and maintainability \cite{ISOIEC5055}. It formalizes metrics for issues such as control flow complexity and code duplication, providing quantitative thresholds for severity.  
\null\quad By systematically identifying code-level weaknesses, ISO/IEC 5055 serves as a critical tool for organizations looking to address deep-seated structural flaws early. In practice, combining static analysis with contextual insights---such as those gleaned from an LLM-based review system---can extend the utility of this standard beyond purely syntactic checks \cite{Pashchenko2021}.

\subsection{ISO 9001 and ISO/IEC 90003}
\label{subsec:iso9001}
Although ISO 9001 applies to quality management systems across industries, its principles of defining processes, documenting procedures, and ensuring continuous improvement underpin software quality assurance \cite{ISO9001}. ISO 9001 mandates risk-based thinking, leadership involvement, and clear documentation of quality objectives.  
\null\quad ISO/IEC 90003 interprets the requirements of ISO 9001 specifically for software development and maintenance \cite{ISOIEC90003}. It provides additional guidance on software-related processes, ensuring organizations align quality management systems with the unique challenges of building and evolving software.

\subsection{Capability Maturity Model Integration (CMMI)}
\label{subsec:cmmi}
Developed by the Software Engineering Institute (SEI), Capability Maturity Model Integration (CMMI) offers a structured approach for process improvement, grouping organizational practices into maturity levels \cite{CMMI2018}. Each level represents a higher degree of process capability, from ad hoc procedures (Level 1) to systematic optimization (Level 5).  
\null\quad CMMI emphasizes quantitative project management and continuous improvement. By integrating LLM-based analytics (e.g., defect prediction, requirement coverage) into CMMI-based practices, organizations can rapidly advance through maturity levels while maintaining robust oversight \cite{Dingsor2019}.

\subsection{Test Maturity Model (TMM)}
\label{subsec:tmm}
The Test Maturity Model (TMM) is designed to assess and improve the maturity of testing processes within software organizations. It delineates stages of test process development, from initial and unstructured to fully optimized \cite{Burnstein2003}. Each level specifies objectives and deliverables that elevate testing quality, such as test planning, monitoring, and control.  
\null\quad TMM’s emphasis on continuous assessment and structured improvements aligns with modern DevOps pipelines. Researchers have highlighted that automated tools, including AI-based test generation, can significantly enhance testing efficiency \cite{Garousi2019}, suggesting that TMM-based organizations could benefit from integrating LLM-driven test automation and analysis.

\section{Large Language Models in Software Engineering}
\label{sec:llm-se}

\subsection{Fundamentals of LLMs}
\label{subsec:fundamentals-llms}

\paragraph{Transformer Architectures.}
The seminal work by Vaswani et al. \cite{Vaswani2017} introduced the transformer model, a cornerstone of contemporary large language models (LLMs). Unlike recurrent neural networks, transformers leverage a self-attention mechanism to capture global dependencies in sequences, allowing models to process text (and code) in parallel rather than sequentially. This design significantly reduces training times and enhances the capacity to learn nuanced relationships from large-scale corpora.

\paragraph{Pre-training and Fine-tuning.}
Modern LLMs undergo a two-stage process: \emph{pre-training} on massive datasets, followed by \emph{fine-tuning} for domain-specific tasks \cite{Devlin2019, Brown2020}. Pre-training typically involves learning general linguistic or code-related representations while fine-tuning refines the model for specialized objectives such as code generation, bug detection, or requirement analysis \cite{Feng2020}.

\paragraph{Emerging Models and Frameworks.}
Recent innovations have produced a variety of transformer-based models tailored for different applications:
\begin{itemize}
    \item \textbf{GPT Series (OpenAI).} In particular, GPT-3 \cite{Brown2020} and GPT-4 focus on natural language generation, boasting billions of parameters trained on diverse internet-scale corpora.
    \item \textbf{BERT Variants (Google).} BERT \cite{Devlin2019}, and its successors (RoBERTa, DistilBERT) use bidirectional context analysis to excel in tasks like sentence classification and question answering.
    \item \textbf{CodeBERT, CodeT5, PolyCoder.} Targeted at code-related tasks, CodeBERT \cite{Feng2020} and CodeT5 \cite{Puri2021} adapt transformer architectures to programming languages, enabling code completion and bug detection. PolyCoder \cite{Xu2022} further explores model architectures optimized for lightweight code understanding and generation.
\end{itemize}

\subsection{Key LLM-based Tasks in Software Engineering}
\label{subsec:llm-tasks-se}

\paragraph{Code Generation and Completion.}
GitHub Copilot \cite{Chen2021} illustrates how advanced LLMs can assist developers by offering real-time suggestions. These suggestions can reduce coding effort, yet they also risk introducing subtle defects if developers become overly reliant on them \cite{Vaithilingam2022}. Balancing productivity gains with thorough code review remains an open challenge.

\paragraph{Requirements Engineering and Analysis.}
LLMs can parse large volumes of textual requirements, highlighting ambiguities or contradictions. Research by Dalpiaz et al. \cite{Dalpiaz2019} demonstrates how semantic analysis can improve requirement clarity, helping teams identify missing or conflicting details early in the development cycle.

\paragraph{Automated Documentation.}
Natural language generation can generate or update documentation based on source code changes. Moreno et al. \cite{Moreno2020} show that although such automation saves significant time, completeness and accuracy persist, underscoring the need for ongoing human oversight and validation.

\paragraph{Test Case Generation.}
LLMs can convert requirement statements or code snippets into test scenarios, enhancing test coverage and catching edge cases. DeepTest \cite{Tian2018} exemplifies how AI-based approaches can automatically craft test cases, notably for complex domains such as autonomous driving software.

\paragraph{Defect Detection and Code Review.}
LLM-based models can flag potential security vulnerabilities, performance bottlenecks, or logical errors when integrated into code review pipelines. By drawing on learned patterns from large code repositories, these models offer a second layer of scrutiny, though explainability and false-positive rates remain concerns.

\paragraph{Compliance and Regulatory Checks.}
Organizations bound by regulations (e.g., HIPAA, GDPR) can leverage LLMs to parse compliance documents and compare them against project artifacts. Discrepancies, such as missing user consent clauses or inadequate data handling procedures, can be highlighted for further investigation.

\subsection{Notable LLM Frameworks and Tools}
\label{subsec:llm-tools}

\paragraph{OpenAI API Ecosystem.}
OpenAI provides endpoints for code generation, text classification, and embedding-based similarity checks. Enterprise-ready features, including role-based access and usage analytics, cater to organizations seeking scalable AI solutions.

\paragraph{Microsoft’s Azure OpenAI Services.}
Microsoft extends OpenAI’s models within Azure, offering compliance-centric features like data encryption at rest, secured endpoints, and traceable audit logs. These services facilitate LLM adoption in sectors with strict regulatory requirements.

\paragraph{Open-Source Tools.}
Beyond commercial offerings, open-source efforts like CodeT5 \cite{Puri2021}, PolyCoder \cite{Xu2022}, and community-driven QA platforms integrate LLM-based bug detection, code completion, and textual analysis. They provide a customizable foundation for teams aiming to build or extend AI-driven quality assurance pipelines without relying on black-box vendor solutions.

\section{Selection Criteria}
\label{selection_criteria}

We established structured selection criteria to ensure our survey remained focused and meaningful at the intersection of Large Language Models (LLMs) and Software Quality Assurance (SQA). We designed these criteria to align with industry-relevant quality goals, demonstrate technical depth, and capture the evolving role of LLMs in software engineering. We included a study in our review if it satisfied the following conditions:

\begin{enumerate}
    \item \textbf{Primary Focus on LLM Use in Software Quality Contexts:} The study investigated the use of LLMs in tasks directly related to software quality assurance, such as requirement analysis, test generation, code quality improvement, bug detection, documentation enhancement, or standards compliance.

    \item \textbf{Alignment with SQA Objectives or Standards:} While explicit references to formal standards (e.g., ISO/IEC 25010, ISO 12207, TMM, or CMMI) were preferred, studies were also included if they demonstrated alignment with core SQA objectives---such as reliability, maintainability, security, test coverage, or suitability ---as defined by these frameworks.

    \item \textbf{Empirical, Technical, or Methodological Rigor:} We prioritized papers that included empirical validation (e.g., case studies, benchmarks, experiments) or detailed methodology (e.g., tool architecture, model design, prompt engineering strategies) to ensure actionable insights and technical credibility.

    \item \textbf{Recency and Technological Relevance:} Considering the rapid advancement of both LLMs and AI-in-SQA tools, we limited our survey to work published from 2023 onward, capturing the post-ChatGPT wave of innovation. We made exceptions for highly cited or foundational papers.

    \item \textbf{Breadth of SQA Activities Represented:} To reflect the diversity of modern SQA, we included papers across multiple quality dimensions---functional correctness, test automation, security auditing, static analysis, configuration validation, and maintainability. This breadth allowed us to evaluate LLM applicability across the full SQA lifecycle.

    \item \textbf{Practical Relevance or Deployment Context:} We preferred studies that demonstrated industrial relevance, provided tool availability, or offered deployment insights (e.g., GitHub Copilot evaluations, Meta’s testing infrastructure) to bridge academic insight with practice.

    \item \textbf{Peer-Reviewed or Preprint Research:} Selected works were drawn from peer-reviewed journals, conference proceedings, or widely recognized preprint repositories (e.g., arXiv). This criterion ensured that all studies met a basic academic rigor and public dissemination standard.

\end{enumerate}

\section{Characteristics of the Surveyed Literature}
To characterize the state of the art in applying large language models (LLMs) to software quality assurance (SQA), we analyzed over 223 papers published between 2023 and 2025. Below, we present key findings drawn from both quantitative distributions and qualitative inspection of representative studies.

\subsection{Publication Trends}
As illustrated in Figure~\ref{fig:papers_per_year}, the number of publications increased sharply after 2023. The field saw 31 papers in 2023, followed by a substantial rise to 91 papers in 2024—nearly a threefold increase. This surge reflects the rapidly growing interest in applying LLMs to SQA, likely driven by the release of more capable open-source and commercial models during this period. In 2025, the count is 127 papers, indicating continued momentum and suggesting that research activity in this area remains strong and accelerating.

\begin{figure}[htbp]
    \centering
    \includegraphics[width=0.8\linewidth]{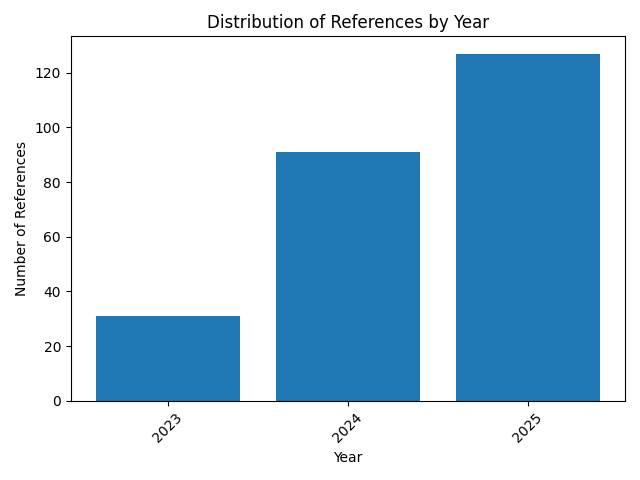}
    \caption{Number of papers published per year from 2023 to 2025, showing consisted growth in interest in LLMs for software quality assurance.}

    \label{fig:papers_per_year}
\end{figure}

\subsection{Dataset Utilization}
\begin{figure}[htbp]
    \centering
    \includegraphics[width=0.9\linewidth]{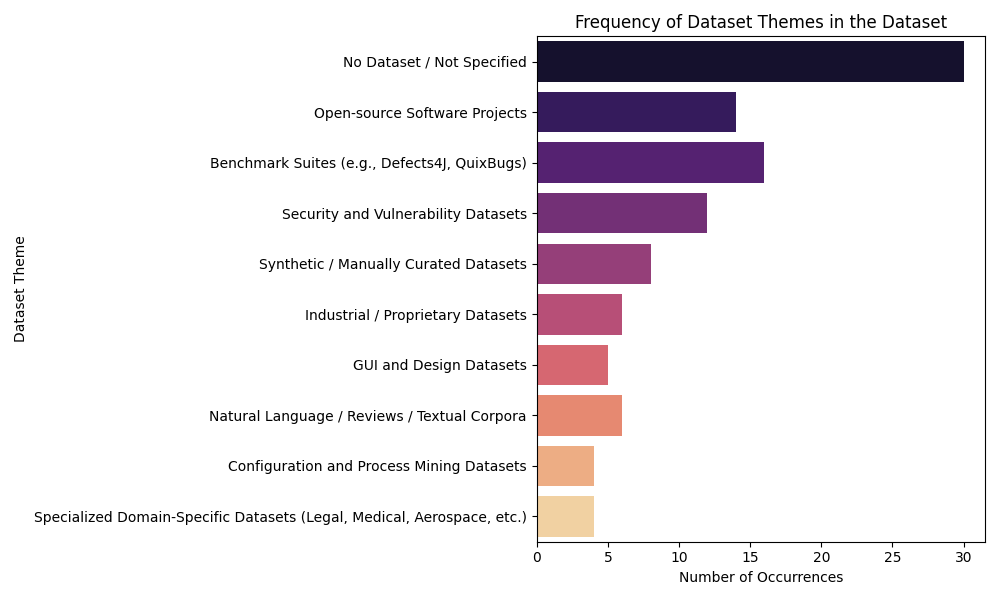}
    \caption{Distribution of dataset themes used in the surveyed literature. A significant portion of papers did not specify any dataset, while open-source projects and benchmark suites were most common among those that did.}

    \label{fig:data_dist}
\end{figure}

Figure~\ref{fig:data_dist} reveals that nearly 30\% of the reviewed papers did not specify any dataset or did not use one at all—highlighting a concerning gap in reproducibility. Among those that did, common dataset types included:
\begin{itemize}
    \item \textbf{Open-source software projects}, such as GitHub issues and repositories.
    \item \textbf{Benchmark suites} like \textit{Defects4J} and \textit{QuixBugs}, particularly popular in testing applications.
    \item \textbf{Security and vulnerability datasets}, as used in works like \textit{LLMs for Intelligent Software Testing: A Comparative Study}.
\end{itemize}
Synthetic and proprietary datasets also appeared but were relatively rare. The use of \textbf{domain-specific datasets} (e.g., medical, legal) remains limited, indicating a focus on general-purpose or software-specific corpora.

\subsection{Evaluation Approaches}
\begin{figure}[htbp]
    \centering
    \includegraphics[width=0.9\linewidth]{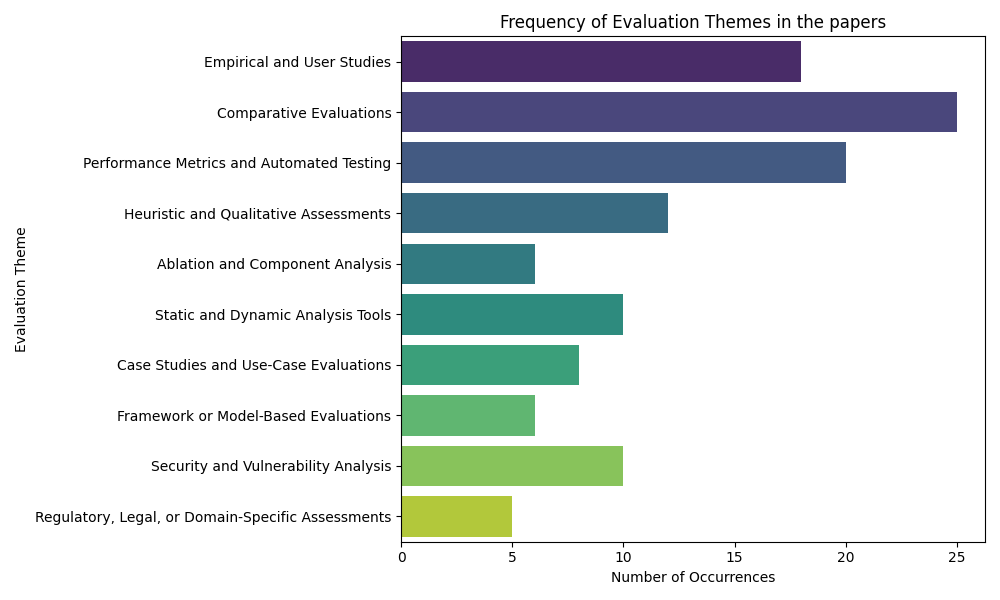}
    \caption{Frequency of evaluation approaches used in the papers. Comparative studies, empirical/user evaluations, and automated performance metrics dominated the landscape.}
    \label{fig:eval_dist}
\end{figure}

Various evaluation methods were employed, as visualized in Figure~\ref{fig:eval_dist}. Common practices included:
\begin{itemize}
    \item \textbf{Comparative evaluations}, e.g., comparing LLM outputs against rule-based or traditional baselines.
    \item \textbf{Empirical/user studies}, as seen in \textit{Advancing Requirements Engineering Through Generative AI}, which collected practitioner feedback.
    \item \textbf{Automated performance metrics}, such as precision, recall, BLEU, and accuracy.
\end{itemize}
A smaller but important subset used ablation analyses, static/dynamic analysis tools, or model-based frameworks.

\subsection{Fine-Tuning Adoption}
\begin{figure}[htbp]
    \centering
    \includegraphics[width=0.5\linewidth]{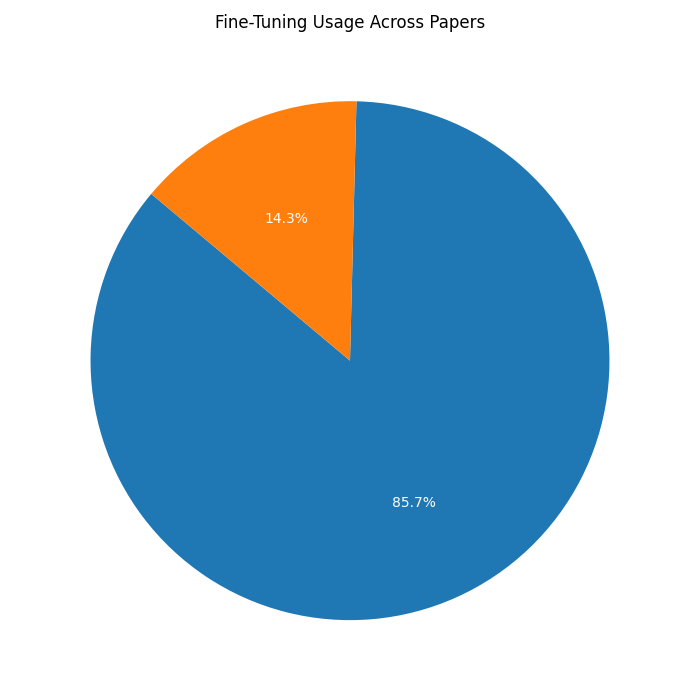}
    \caption{Proportion of papers that used fine-tuned LLMs versus those that relied solely on pre-trained models. Most studies avoided fine-tuning.}

    \label{fig:finetuned}
\end{figure}

As shown in Figure~\ref{fig:finetuned}, only 14.3\% of the studies reported using fine-tuned models. Most leveraged zero-shot or few-shot prompting on pre-trained models like ChatGPT or GPT-4. One exception is \textit{Requirements are All You Need: From Requirements to Code with LLMs}, which fine-tunes on domain-specific data to improve generation accuracy.

\subsection{LLMs in Use}
\begin{figure}[htbp]
    \centering
    \includegraphics[width=0.8\linewidth]{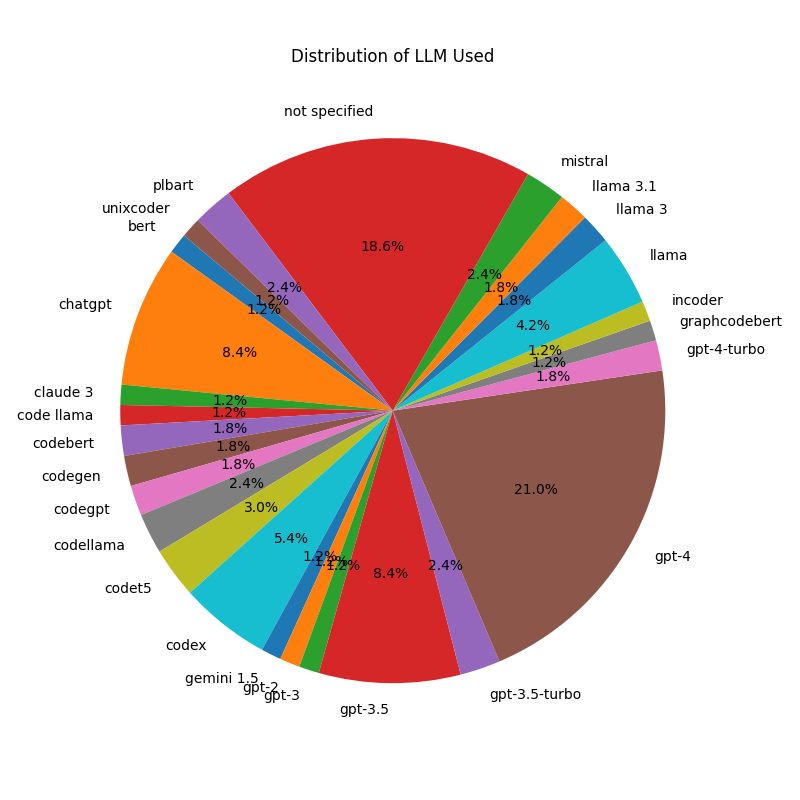}
    \caption{Distribution of LLMs reported in the literature. GPT-4, GPT-3.5, and ChatGPT were the most commonly used, though many papers did not specify the model used.}
    \label{fig:llms_used}
\end{figure}

Figure~\ref{fig:llms_used} shows that GPT-4 was the most commonly used model (21\%), followed by GPT-3.5, ChatGPT, and CodeT5. Open-source models like LLaMA and CodeLLaMA were used in academic or experimental studies. Nearly 19\% of papers did not specify which LLM was used, raising reproducibility concerns.

\subsection{Prompting Strategies}
\begin{figure}[htbp]
    \centering
    \includegraphics[width=0.8\linewidth]{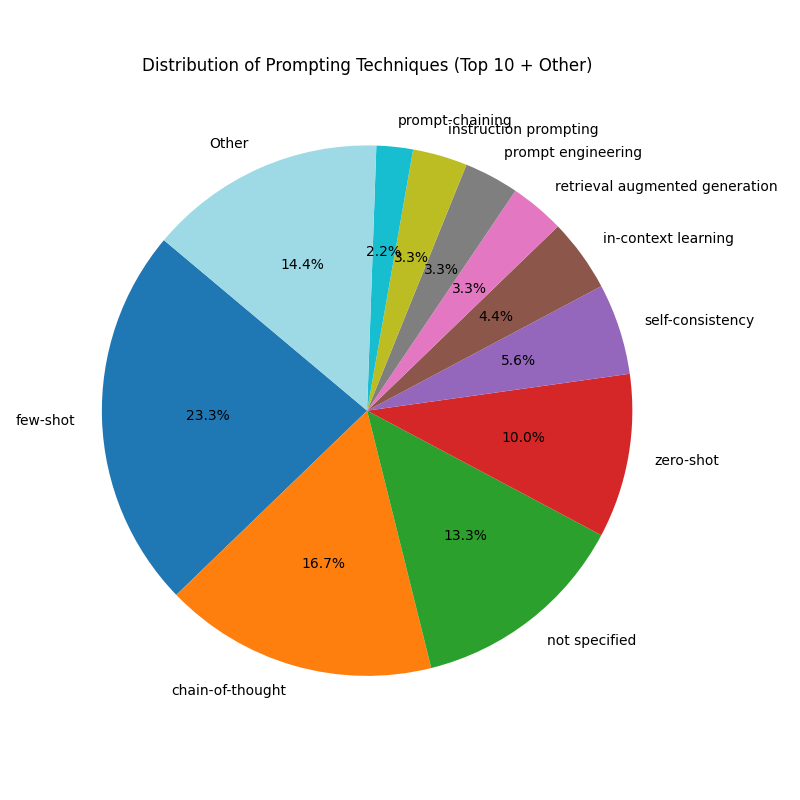}
    \caption{Distribution of prompting techniques employed across papers. Few-shot and chain-of-thought prompting were most prevalent, with some papers using custom or hybrid strategies.}
    \label{fig:prompts}
\end{figure}

Prompting strategies varied significantly (see Figure~\ref{fig:prompts}). Few-shot prompting was most popular (23.3\%), followed by chain-of-thought (16.7\%) and zero-shot (10\%). A notable portion (14.4\%) used ``other'' techniques, covering custom or hybrid strategies. More advanced techniques, such as instruction prompting, prompt chaining, and retrieval-augmented generation, were occasionally employed.

In summary, the literature on LLMs in SQA is expanding rapidly. Most studies rely on general-purpose LLMs and prompt-based interaction rather than fine-tuning. Evaluation practices are becoming more diverse and rigorous, though dataset usage and reporting gaps remain.

\begin{table*}[htbp]
\centering
\caption{Extended LLM-SQA Standards Mapping (Per Application)}
\label{tab:detailed}
\begin{tabularx}{\textwidth}{l l Y Y}
  \toprule
  \textbf{Standard} & 
  \textbf{Quality Attribute/Process} & 
  \textbf{LLM Application} & 
  \textbf{Key References} \\
  \midrule

  \multirow{6}{*}{ISO/IEC 12207} 
  & Software Requirements & Translate stakeholder needs into requirement specifications & \cite{krishna2024using, wei2024requirements, arora2024advancing, lutze2024generating, lubos2024leveraging, white2024chatgpt, couder2024requirements,ocleppo2025enhancing,robinson2025requirements,voria2025recover,ge2025cross,kong2025collaboration,vogelsang2025using,alhaizaey2025automated,guo2025natural} \\
  & Architecture and Design & Recommend modular architecture and design patterns & \cite{white2024chatgpt, xu2025mantra, jelodar2025large, cordeiro2024empirical, arora2024advancing, tagliaferro2025leveraging,pandey2025design,gu2025large,zhang2025knowledge,guerra2025assessing,daareyni2025generative,cheung2025composing,vungarala2025sa} \\
  & Implementation & Generate code templates and enforce standards & \cite{fakhoury2024llm, rasheed2024ai, pomian2024assist, pomian2024furtherllmsidestatic, wu2024ismell, xu2025mantra, jelodar2025large, collini2025c2hlsc,huang2025template,huang2025opencoder,gu2025effectiveness,peng2025soleval,tian2025fixing,weyssow2025exploring,zhang2025scalable} \\
  & Verification & Generate unit and integration tests & \cite{alagarsamy2024enhancing, arora2024generating, li2025evaluating, sami2024tool, guilherme2023initial, bhatia2024system, fakhoury2024llm, pan2024multi, foster2025mutation, cai2025automated,zhang2025exploring,alkafaween2025automating,sun2025classinvgen,shang2025large,lops2025system,wang2025testeval,biagiola2025improving,cheng2025rug,ouedraogo2025enriching,kim2025llamaresttest} \\
  & Validation & Compare test outcomes against requirements & \cite{foster2025mutation, rahman2024automated, bhatia2024system, alagarsamy2024enhancing, arora2024generating, zhang2025patch,huang2025comprehensive,shah2025using} \\
  & Configuration Management & Track revisions and suggest updates & \cite{lian2024large, lian2023configuration, wen2024llm, wang2024identifying, pornprasit2024fine, shan2024face, wang2024netconfeval, lian2025large,tu2025intent,chen2025automatic,giannakouris2025lambda,angi2025llnet,ye2025llmsecconfig,wen2025llm} \\
  \midrule
  
  \multirow{8}{*}{ISO/IEC 25010} 
  & Functional Suitability & Generate requirement-aligned test cases & \cite{krishna2024using, wei2024requirements, arora2024generating, alagarsamy2024enhancing, lutze2024generating, lubos2024leveraging, couder2024requirements, rahman2024automated, bhatia2024system, sami2024tool, pan2024multi, guilherme2023initial, de2025comparison} \\
  & Reliability & Detect unstable patterns and improve fault tolerance & \cite{fakhoury2024llm, foster2025mutation, hu2024use, pelliccione2024insights, alshahwan2024assured, purba2023software, andrade2025enhancing,jin2025adaptive,salman2025hybrid,ji2025cloud,xu2025openrca} \\
  & Usability & Review and enhance UI/UX clarity and accessibility & \cite{duan2024generating, petridis2023promptinfuser, duan2024uicrit, duan2023towards, brie2023evaluating, kolthoff2024interlinking, lu2025uxagent, ghosh2024enhancing, zhang2024exploring, sun2024llms, bassi2025supporting,fang2025enhancing,yuan2025towards,park2025leveraging,ahmed2025multimodal,li2025portal,iman2025refining} \\
  & Performance Efficiency & Optimize inefficient code constructs & \cite{wadhwa2024core, chow2024performance, wei2025improvingparallelprogramperformance, gao2024search, cui2025large, rosas2024should, coignion2024performance, niu2024evaluating, ye2025llm4effi,purschke2025speedgen,italiano2025finding,almatrafi2025code} \\
  & Maintainability & Recommend refactorings and reduce code smells & \cite{xu2025mantra, cordeiro2024empirical, electronics13091644, pomian2024furtherllmsidestatic, ZHANG2025121753, wu2024ismell, pomian2024assist, jelodar2025large, nunes2025evaluating, liu2025exploring,liang2025smelldetector,asvydyte2025well,pandini2025exploratory,tornhill2025ace,xue2025clean,liu2024empirical,alomar2024refactor} \\
  & Portability & Ensure cross-platform compatibility & \cite{wang2024identifying, albuquerque2024evaluating, lian2024large, wen2024llm, lian2023configuration, wang2024netconfeval, muneer2025meta} \\
  & Compatibility & Analyze integration and dependency issues & \cite{rasheed2024ai, lian2024large, lu2023llama, zhang2024detecting, pornprasit2024fine, shan2024face, malmqvist2025enhancing,giabbanelli2025over,bartlett2025raiders} \\
  & Security & Flag insecure practices and recommend secure alternatives & \cite{li2024llm, zibaeirad2025reasoning, yang2025code, yang2025context, akuthota2023vulnerability, islam2024llm, ullah2024llms, saha2024llm, guo2024outside, kulsum2024case, wu2023effective, boi2024smart, noever2023can, pearce2023examining, de2024enhanced, sajadi2025llms,andrade2025enhancing,chennabasappa2025llamafirewall} \\
  \midrule

  \multirow{4}{*}{ISO/IEC 5055} 
  & Reliability (Code-Level) & Detect error-prone constructs, support fault localization & \cite{foster2025mutation, hu2024use, pelliccione2024insights, arora2024generating, bhatia2024system, fakhoury2024llm, pan2024multi, pu2025errorprism,xu2025two,xu2025openrca,sovrano2025large} \\
  & Performance (Code-Level) & Detect inefficiencies, optimize performance & \cite{wadhwa2024core, chow2024performance, gao2024search, coignion2024performance, cui2025large, wei2025improvingparallelprogramperformance, niu2024evaluating, rosas2024should, wadhwa2023frustrated, huang2025autonomous,almatrafi2025code,ye2025llm4effi} \\
  & Security (Code-Level) & Security audits, vulnerability detection & \cite{li2024llm, zibaeirad2025reasoning, yang2025code, yang2025context, islam2024llm, akuthota2023vulnerability, guo2024outside, kulsum2024case, ullah2024llms, pearce2023examining, wu2023effective, noever2023can, boi2024smart, de2024enhanced, khare2025understanding,tamberg2025harnessing,shestov2025finetuning,tihanyi2025new,liu2025enhancing} \\
  & Maintainability (Code-Level) & Detect low maintainability traits, suggest refactoring & \cite{xu2025mantra, cordeiro2024empirical, electronics13091644, pomian2024furtherllmsidestatic, ZHANG2025121753, wu2024ismell, pomian2024assist, jelodar2025large, nunes2025evaluating, gogani2025technical,rong2025code,alturayeif2025refactoring,chand2025automated} \\
  \midrule

  \multirow{7}{*}{ISO 9001/9003} 
  & Customer Focus & Extract insights from customer feedback & \cite{mukku2024insightnet, zhang2024allhands, lin2024interpretable, wang2024llm, falatouri2024harnessing, soni2023large, wulf2024exploringpotentiallargelanguage, agua2025large,wei2025user,bai2025analysis,yu2025application,roumeliotis2025think,kyriakidis2025extracting} \\
  & Leadership & Draft policies and quality objectives & \cite{zhu2024llm, berti2023processmining, kourani2024process, mandvikar2023augmenting, fukuda2025development,kourani2025evaluating} \\
  & People Engagement & Generate onboarding and training content & \cite{pereira2024leveraging, kernan2024knowledge} \\
  & Process Approach & Standardize process documentation & \cite{zhu2024llm, berti2023processmining, kourani2024process, mandvikar2023augmenting, fukuda2025development,kourani2025evaluating} \\
  & Continuous Improvement & Analyze historical data and recommend refinements & \cite{berti2024evaluating, berti2023leveraginglargelanguagemodels, su2023hotgpt} \\
  & Evidence-Based Decisions & Summarize metrics and logs & \cite{godbole2024leveraging, leivaaraos2025large, chen2024data, zhang2024largelanguagemodelsdata, jansen2025leveraging, ding2024data, sui2024table, nasseri2023applications} \\
  & Relationship Management & Analyze communication logs & \cite{liu2023voicesvalidityleveraginglarge, calderon2025behalfstakeholderstrendsnlp, huang2025crmarena} \\
  \midrule

  \multirow{4}{*}{CMMI 2.0} 
  & Governance & Support audits and risk assessments & \cite{arora2024advancing, dzeparoska2023llm, feng2024policy, cheong2024not, arora2024towards, sovrano2025simplifying, guldimann2024compl, li2025llmes} \\
  & Operations & Analyze logs and process inefficiencies & \cite{shan2024face, wen2024llm, lu2023llama, rasheed2024ai, lian2024large, wang2024identifying, pornprasit2024fine, albuquerque2024evaluating, zhang2024detecting, galdino2025large} \\
  & Support & Create SOPs and internal guides & \cite{pereira2024leveraging, kernan2024knowledge, naimi2024automating, della2024using, berti2023processmining, xu2025mantra, wu2024ismell, pomian2024assist, wang2025sop,karacapilidis2025ai} \\
  & Strategic Planning & Summarize trends and support forecasting & \cite{soru2024trend, alzapiedi2024trend, godbole2024leveraging, leivaaraos2025large, zhang2024largelanguagemodelsdata, jansen2025leveraging, calderon2025behalfstakeholderstrendsnlp, su2023hotgpt} \\
  \midrule

  \multirow{5}{*}{TMM} 
  & Level 1 - Initial & Generate baseline unit tests & \cite{guilherme2023initial, li2025evaluating, boukhlif2024llms, santos2024we, bayri2023ai} \\
  & Level 2 - Defined & Create structured test plans and traceability & \cite{krishna2024using, wei2024requirements, arora2024generating, lubos2024leveraging, sami2024tool, vogelsang2025impact} \\
  & Level 3 - Integrated & Integrate tests with CI/CD pipelines & \cite{fakhoury2024llm, pan2024multi, rahman2024automated, couder2024requirements, foster2025mutation} \\
  & Level 4 - Managed & Analyze test metrics and coverage & \cite{alagarsamy2024enhancing, bhatia2024system, santos2024we, li2025evaluating, arora2024generating} \\
  & Level 5 - Optimized & Recommend test optimization strategies & \cite{foster2025mutation, fakhoury2024llm, sami2024tool, bayri2023ai, 10962454, 10.1145/3639478.3643119} \\
  \bottomrule
\end{tabularx}
\label{tab:llm-standard-applications}
\end{table*}

\section{Practical Implementation Blueprint for LLM-Enhanced SQA}

While prior sections explored the conceptual and empirical alignment between Large Language Models (LLMs) and Software Quality Assurance (SQA) standards, practitioners require actionable guidance for adoption. This section presents a step-by-step implementation blueprint to integrate LLM-based techniques into existing SQA processes in a controlled, standards-aligned manner.

\subsection{Step 1: Assess Organizational SQA Maturity}

Organizations should begin by evaluating their current SQA maturity using established frameworks such as ISO/IEC 12207, CMMI, or the Test Maturity Model (TMM). This assessment helps identify process gaps and determines where LLM-based augmentation can provide the highest impact. For example, organizations at lower maturity levels may benefit from automated test generation, whereas higher-maturity organizations can leverage predictive analytics and compliance automation.

\subsection{Step 2: Identify High-Impact Use Cases}

Based on the maturity assessment, teams should map key pain points in the software lifecycle to LLM-supported tasks. Common high-value use cases include:
\begin{itemize}
    \item Requirement ambiguity detection and validation
    \item Automated test case generation and coverage enhancement
    \item Code review and defect detection
    \item Documentation generation and maintenance
    \item Compliance verification against regulatory standards
\end{itemize}

\subsection{Step 3: Select Deployment Model}

Organizations must choose an appropriate deployment strategy based on security, scalability, and compliance requirements:
\begin{itemize}
    \item \textbf{Public API-based models:} to deploy, suitable for non-sensitive data
    \item \textbf{Private cloud deployments:} balanced control and scalability
    \item \textbf{On-premises models:} maximum data security and regulatory compliance
\end{itemize}

The choice should align with data governance policies and regulatory constraints (e.g., GDPR, HIPAA).

\subsection{Step 4: Integrate into Existing Toolchains}

LLM capabilities should be embedded within existing development and DevOps pipelines rather than treated as standalone tools. Integration points include:
\begin{itemize}
    \item IDE plugins for real-time code assistance
    \item CI/CD pipelines for automated testing and validation
    \item Issue tracking systems for requirement and defect analysis
    \item Code repositories for continuous quality monitoring
\end{itemize}

\subsection{Step 5: Establish Governance and Oversight}

To ensure trustworthiness and compliance, organizations must implement governance mechanisms:
\begin{itemize}
    \item Logging and traceability of LLM-generated outputs
    \item Human-in-the-loop validation for critical decisions
    \item Periodic audits for bias, accuracy, and compliance
    \item Documentation of model usage and limitations
\end{itemize}

These practices align with ISO 9001 principles of accountability and continuous improvement.

\subsection{Step 6: Define Evaluation Metrics}

Quantitative metrics should be established to measure the effectiveness of LLM integration:
\begin{itemize}
    \item Defect detection rate
    \item Test coverage improvement
    \item Reduction in manual review effort
    \item False positive/negative rates
    \item Compliance adherence metrics
\end{itemize}

Baseline comparisons against traditional methods are recommended to validate performance gains.

\subsection{Step 7: Pilot and Scale}

Organizations should initiate adoption through controlled pilot projects targeting a single phase of the software lifecycle (e.g., testing or requirements analysis). Based on pilot results:
\begin{itemize}
    \item Refine prompt engineering and workflows
    \item Expand to additional lifecycle stages
    \item Introduce fine-tuning or domain adaptation if necessary
\end{itemize}

Gradual scaling ensures risk mitigation and sustainable adoption.

\subsection{Step 8: Continuous Monitoring and Improvement}

Given the evolving nature of both software systems and LLM capabilities, continuous monitoring is essential:
\begin{itemize}
    \item Track model performance over time
    \item Update models with new data (incremental fine-tuning)
    \item Incorporate developer feedback (active learning)
    \item Adapt to evolving standards and regulatory requirements
\end{itemize}

This iterative approach aligns with continuous improvement practices in mature SQA frameworks.

\subsection{Summary}

This implementation blueprint provides a structured pathway for integrating LLMs into SQA processes. By aligning adoption with established standards, governance practices, and measurable outcomes, organizations can realize the benefits of automation while maintaining reliability, compliance, and accountability.

\begin{figure*}[htbp]
    \centering
    \includegraphics[width=0.8\linewidth]{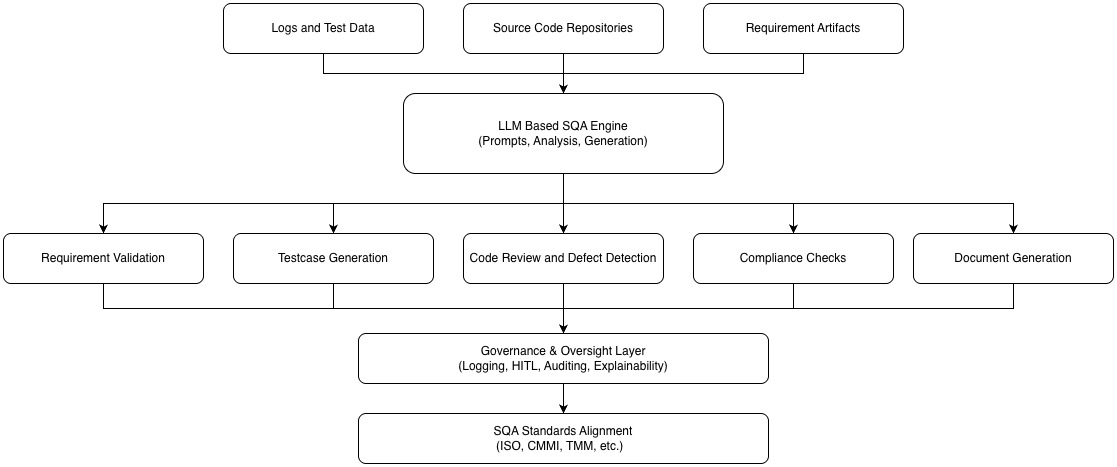}
    \caption{Proposed Architecture of LLM-Enhanced Software Quality Assurance (SQA) Framework. The diagram illustrates how LLM-based components integrate with software artifacts, generate quality assurance outputs, and operate under governance mechanisms aligned with established standards.}
    \label{fig:papers_per_year}
\end{figure*}

\section{Decision Matrix for LLM Adoption in SQA}

To support practitioners in selecting appropriate LLM-based interventions, we present a decision matrix that maps common software quality objectives to corresponding LLM applications, associated standards, and relative risk levels. This matrix serves as a practical guide for prioritizing adoption based on organizational needs.

\subsection{LLM Adoption Decision Matrix}

\begin{table*}[h]
\centering
\caption{Decision Matrix for LLM-Based SQA Adoption}
\begin{tabular}{|p{3.5cm}|p{4cm}|p{3.5cm}|p{2cm}|}
\hline
\textbf{Quality Objective} & \textbf{Recommended LLM Application} & \textbf{Relevant Standards} & \textbf{Risk Level} \\
\hline
Improve requirement clarity & Ambiguity detection, requirement validation & ISO/IEC 12207, ISO/IEC 25010 & Low \\
\hline
Enhance test coverage & Automated test case generation, edge-case synthesis & TMM, ISO/IEC 25010 & Low \\
\hline
Increase code reliability & AI-assisted code review, defect detection & ISO/IEC 5055, ISO/IEC 25010 & Medium \\
\hline
Strengthen security posture & Vulnerability detection, secure coding recommendations & ISO/IEC 5055, ISO/IEC 25010 & Medium \\
\hline
Improve maintainability & Code refactoring suggestions, code smell detection & ISO/IEC 25010, ISO/IEC 5055 & Low \\
\hline
Optimize performance & Detection of inefficient code patterns & ISO/IEC 5055 & Medium \\
\hline
Automate documentation & Code-to-documentation generation and updates & ISO 9001/90003 & Low \\
\hline
Support compliance auditing & Policy analysis, requirement-to-regulation mapping & ISO 9001, CMMI & Medium \\
\hline
Improve process efficiency & Log analysis, workflow optimization & CMMI & Medium \\
\hline
Enable continuous improvement & Trend analysis, defect prediction, feedback summarization & ISO 9001, CMMI & Medium \\
\hline
\end{tabular}
\end{table*}

\subsection{Interpretation Guidelines}

The decision matrix can be used as follows:
\begin{itemize}
    \item \textbf{Low-risk applications} (e.g., test generation, documentation) are suitable for early adoption and pilot projects.
    \item \textbf{Medium-risk applications} (e.g., security analysis, compliance checks) require stronger governance, validation, and human oversight.
    \item Organizations should prioritize use cases aligned with their most critical quality objectives and maturity level.
\end{itemize}

\subsection{Strategic Use}

The matrix enables:
\begin{itemize}
    \item Rapid identification of high-impact LLM use cases
    \item Alignment of AI adoption with recognized SQA standards
    \item Risk-aware planning for incremental deployment
\end{itemize}

By combining this matrix with the implementation blueprint in the previous section, organizations can transition from exploratory experimentation to structured, standards-aligned adoption of LLM-based SQA techniques.

\section{Challenges, Limitations, and Risks}
\label{sec:challenges}

\subsection{Data Privacy and Security}
\label{subsec:data-privacy}

\paragraph{Risk of Data Exposure}
One of the most pressing concerns is the potential exposure of proprietary or sensitive code when using public APIs. Such unintentional leaks could violate confidentiality agreements or data protection mandates \cite{CloudSecurityAlliance2019}. Sensitive information embedded in code, such as API keys or personal data, may be inadvertently processed by third-party services.

\paragraph{Mitigation Strategies}
Organizations can opt for on-premises deployment or secure private cloud solutions to retain internal control over data flow. Additionally, fine-tuning models on anonymized or obfuscated datasets reduces the risk of revealing sensitive information while allowing domain-specific improvements. This trade-off between privacy and model performance highlights the need for robust data governance frameworks \cite{GDPR2018}.

\subsection{Model Bias and Ethical Considerations}
\label{subsec:model-bias}

\paragraph{Training Data Bias}
LLMs trained on large, uncurated datasets frequently inherit biases reflected in their training corpora \cite{Mehrabi2021}. In software contexts, such biases might manifest as limited coverage of diverse user requirements, or preferential treatment of certain coding patterns prevalent in the training data.

\paragraph{Ethical and Legal Implications}
Automated quality assurance decisions can overlook domain-specific ethical or legal requirements, such as patient confidentiality in healthcare applications \cite{Gama2020}. Over-reliance on LLM outputs could lead to decisions that fail to account for contextual nuances or local regulations. Consequently, human oversight remains essential to prevent ethically or legally problematic outcomes.

\subsection{Explainability and Transparency}
\label{subsec:explainability}

\paragraph{Black Box Concerns}
Transformers and other large neural architectures often lack inherent interpretability, making it challenging to justify the reasoning behind a particular suggestion \cite{Rudin2019}. In high-stakes domains like finance, aerospace, or healthcare, stakeholders may be unwilling to adopt AI-driven QA without clearer insights into how recommendations are generated.

\paragraph{Potential Solutions}
Research into explainable AI (XAI) seeks to bridge this gap by proposing techniques such as attention visualization, summarized reasoning chains, or supplementary symbolic analysis \cite{DoshiVelez2017}. Although these methods add transparency, they can increase computational overhead and complexity, highlighting a tension between model performance and interpretability.

\subsection{Resource Requirements and Scalability}
\label{subsec:resources}

\paragraph{Computational Demands}
Many state-of-the-art LLMs boast tens or hundreds of billions of parameters, necessitating substantial processing power and memory. Running these models at scale can incur high hardware, energy, and maintenance costs- a barrier for small to medium-sized enterprises with limited IT budgets \cite{Shaffer2021}.

\paragraph{Cost--Benefit Analysis}
Organizations must weigh the potential gains in QA productivity and defect reduction against the expenses of acquiring and maintaining the requisite infrastructure. Some teams adopt a hybrid approach, using lighter, distilled models for routine tasks while reserving larger, more expensive models for complex or mission-critical analyses.

\subsection{Governance and Auditing}
\label{subsec:governance}

\paragraph{Traceability and Logging}
Robust governance frameworks require detailed logs of AI-driven decisions and developer overrides, ensuring accountability in scenarios where LLM-suggested changes lead to production issues \cite{Raji2020}. These logs also facilitate post-deployment reviews and retrospective analyses, essential for regulated industries like finance or healthcare.

\paragraph{Standards Compliance}
Many industry-specific regulations mandate auditable processes for software changes \cite{ISO9001}. As organizations integrate LLMs into QA pipelines, they must demonstrate compliance with relevant ISO standards, legal statutes, and sector-specific best practices. This necessitates thorough documentation of how models are trained, validated, and continuously monitored.

\section{Future Directions}
\label{sec:future-directions}

The convergence of Large Language Models (LLMs) and Software Quality Assurance (SQA) has already demonstrated significant potential. However, further innovations and research areas remain ripe for exploration. This section highlights emerging directions in model adaptability, deployment architectures, multimodal analysis, standards evolution, and cross-industry applications.

\subsection{Adaptive and Continual Learning}
\label{subsec:adaptive-learning}

\paragraph{Incremental Fine-tuning}
As codebases evolve, static models may become outdated, missing recent changes or new coding conventions. A promising direction is \emph{incremental fine-tuning}, where LLMs continuously learn from recent commits, bug reports, and updated libraries \cite{Sun2019}. Such adaptability can preserve or even improve accuracy over time, provided organizations balance retraining costs with tangible improvements in defect detection and code generation.

\paragraph{Active Learning Paradigms}
Active learning leverages human intervention to refine the model. For instance, when an LLM detects ambiguous or high-risk outputs, it can query developers for clarification. These labeled examples feed back into the model, enhancing its understanding of domain-specific patterns \cite{Settles2009}. This iterative feedback loop aligns well with agile development cycles, enabling more responsive and context-aware QA.

\subsection{Federated or On-Premises LLM Deployments}
\label{subsec:federated-onprem}

\paragraph{Privacy-Focused Approaches}
Federated learning offers a method to aggregate insights from multiple sources without transferring raw data to a centralized server \cite{Kairouz2019}. Only model parameters are shared in this model, not the underlying code or sensitive information. This design can help meet stringent regulatory requirements in domains like healthcare or finance.

\paragraph{Industry Examples}
Privacy-conscious sectors like banking have begun piloting secure, private AI deployments \cite{Emam2020}. By conducting model training and inference on-premises or within tightly controlled cloud environments, organizations maintain compliance with confidentiality and data protection regulations while still benefiting from advanced LLM-based QA.

\subsection{Advanced Multimodal and Integrated SQA}
\label{subsec:multimodal-integration}

\paragraph{Beyond Text and Code}
While current SQA practices often focus on source code and natural language requirements, real-world systems encompass multiple data modalities. Future approaches may integrate UML diagrams, real-time performance metrics, and user feedback logs, providing a more holistic view of software quality \cite{Hecht2019}.

\paragraph{Potential Gains}
Multimodal analysis can detect complex architectural flaws or emergent behaviors that text- or code-only methods might miss. For instance, a spike in user-reported issues alongside an unusual surge in memory usage could indicate a regression in system design. By correlating these signals, LLM-driven QA tools could propose deeper, more targeted improvements earlier in the development life cycle.

\subsection{Standard Updates to Incorporate AI/LLM Methodologies}
\label{subsec:standard-updates}

\paragraph{Call for AI-Specific Guidelines}
Existing SQA standards, such as ISO/IEC 25010 and ISO/IEC 5055, provide robust frameworks for measuring software quality but do not explicitly address AI-based tooling. As LLMs permeate more QA processes, there is a growing need for dedicated guidelines or amendments to outline best practices for AI-based testing, validation, and compliance \cite{Wang2021}.

\paragraph{Regulatory Initiatives}
Governmental and international bodies are increasingly aware of the impact of AI on safety and ethics. For instance, the proposed EU AI Act would require transparent documentation and rigorous risk assessments for high-stakes AI systems \cite{EuropeanCommission2021}. Incorporating these mandates into software quality standards will be critical to ensuring both innovation and accountability.

\subsection{Cross-Industry and Cross-Domain Extensions}
\label{subsec:cross-industry}

\paragraph{Knowledge Transfer}
Techniques pioneered in one sector---such as automotive or aerospace---can be adapted for others, accelerating AI adoption in less digitized industries \cite{Pan2010}. For example, diagnostic methods that detect sensor data anomalies in autonomous vehicles could inform QA practices in industrial robotics or medical devices.

\paragraph{Open Research Questions}
Many open questions remain about handling domain-specific terminologies, compliance norms, and testing workflows across different fields. Further investigation is needed to develop generalized methods that retain enough adaptability to suit varied regulatory landscapes and stakeholder requirements, enabling LLM solutions to scale seamlessly across diverse domains.

\bibliographystyle{IEEEtran}
\bibliography{ref}
\vspace{12pt}

\end{document}